\documentclass[acmtog, review=false]{acmart}

\usepackage{amsmath}
\usepackage{bm}
\usepackage{algorithm}
\usepackage[noend]{algpseudocode}
\usepackage{booktabs}
\usepackage{amssymb}
\usepackage{hyperref}
\usepackage{multirow}
\usepackage{tabu}
\usepackage{tabularx}
\usepackage{tabulary}
\usepackage{ragged2e}
\usepackage{xspace}
\usepackage{subcaption}
\usepackage{paralist}
\usepackage{tabularx}
\usepackage{tabu}
\usepackage{rotating}
\usepackage{layouts}
\usepackage{microtype}
\usepackage{lineno}
\usepackage{soul}
\usepackage{mathtools}
\usepackage{bbm}
\usepackage{soul}
\usepackage{etoolbox}
\usepackage[group-separator={,}]{siunitx}
\usepackage{wrapfig,lipsum,booktabs}
\usepackage{afterpage}
\robustify\bfseries

\graphicspath{{images/}}

\usepackage{ulem}
\pdfpageattr {/Group << /S /Transparency /I true /CS /DeviceRGB>>}

\newcommand{\abs}[1]{\left\lvert #1 \right\rvert}
\newcommand{\norm}[1]{\left\lVert #1 \right\rVert}

\newcommand{\ADD}[1]{#1}
\newcommand{\TODO}[1]{}
\newcommand{\RM}[1]{}

\newcolumntype{Y}{>{\RaggedRight\arraybackslash}X}

\makeatletter
\def\and{
  \end{tabular}%
  \hskip .3em \@plus.17fil%
  \begin{tabular}[t]{c}}
\makeatother

\setcitestyle{square}

\newcommand{\notation}[1]{\ensuremath{#1}\xspace}

\newcommand{\na}{{--}}
\newcommand*{\eg}{e.g.\@\xspace}
\newcommand*{\ie}{i.e.\@\xspace}

\let\originalleft\left
\let\originalright\right
\renewcommand{\left}{\mathopen{}\mathclose\bgroup\originalleft}
\renewcommand{\right}{\aftergroup\egroup\originalright}

\newcommand{\Input}{\notation{\textup{I}}}
\newcommand{\DownsampledInput}{\notation{\tilde{\textup{I}}}}
\newcommand{\Output}{\notation{\textup{O}}}
\newcommand{\Guide}{\notation{g}}

\newcommand{\InputChannels}{\notation{3}}

\newcommand{\GridDepth}{\notation{d}}

\newcommand{\tent}{\notation{\tau}}
\newcommand{\WidthRatio}{\notation{s_x}}
\newcommand{\HeightRatio}{\notation{s_y}}

\newcommand{\CurveThreshold}{\notation{t}}
\newcommand{\CurveSlope}{\notation{a}}
\newcommand{\Curve}{\notation{\rho}}
\newcommand{\ColorCorrectionMatrix}{\notation{\bm{M}}}
\newcommand{\NCurvePoints}{\notation{16}}

\newcommand{\lab}{\notation{\textup{L*a*b*}}}

\newcommand{\dataset}{\notation{\mathcal{D}}}
\newcommand{\learningrate}{\notation{10^{-4}}}
\newcommand{\weightdecay}{\notation{10^{-8}}}
\newcommand{\Loss}{\notation{\mathcal{L}}}
\newcommand{\Weight}{\notation{w}}
\newcommand{\Bias}{\notation{b}}
\newcommand{\Activation}{\notation{\sigma}}
\newcommand{\Shared}{\notation{S}}

\newcommand{\Global}{\notation{G}}
\newcommand{\Local}{\notation{L}}
\newcommand{\NShared}{\notation{n_\Shared}}
\newcommand{\NGlobal}{\notation{n_\Global}}
\newcommand{\NLocal}{\notation{n_\Local}}
\newcommand{\Fused}{\notation{F}}
\newcommand{\ParameterLayer}{\notation{A}}
\newcommand{\SlicedParameterLayer}{\notation{\bar{\ParameterLayer}}}
\newcommand{\Stride}{\notation{s}}
\newcommand{\FullresFeatures}{\notation{\phi}}
\newcommand{\NFullresFeatures}{\notation{n_\FullresFeatures}}

\newcommand{\hz}{\notation{\,\textup{Hz}}}
\newcommand{\ms}{\notation{\,\textup{ms}}}
\newcommand{\secs}{\notation{\,\textup{s}}}

\title{Deep Bilateral Learning for Real-Time Image Enhancement}

\author{Micha\"el Gharbi} \affiliation{\institution{MIT CSAIL}} \email{gharbi@mit.edu}
\author{Jiawen Chen} \affiliation{\institution{Google Research}} \email{jiawen@google.com}
\author{Jonathan T. Barron} \affiliation{\institution{Google Research}} \email{barron@google.com}
\author{Samuel W. Hasinoff} \affiliation{\institution{Google Research}} \email{hasinoff@google.com}
\author{Fr\'edo Durand} \affiliation{\institution{MIT CSAIL / Inria,
Universit\'e C\^ote d'Azur}} \email{fredo@mit.edu}

\keywords{real-time image processing, deep learning, data-driven methods,
convolutional neural networks}

\begin{document}

\begin{teaserfigure}
  \includegraphics[width=\textwidth]{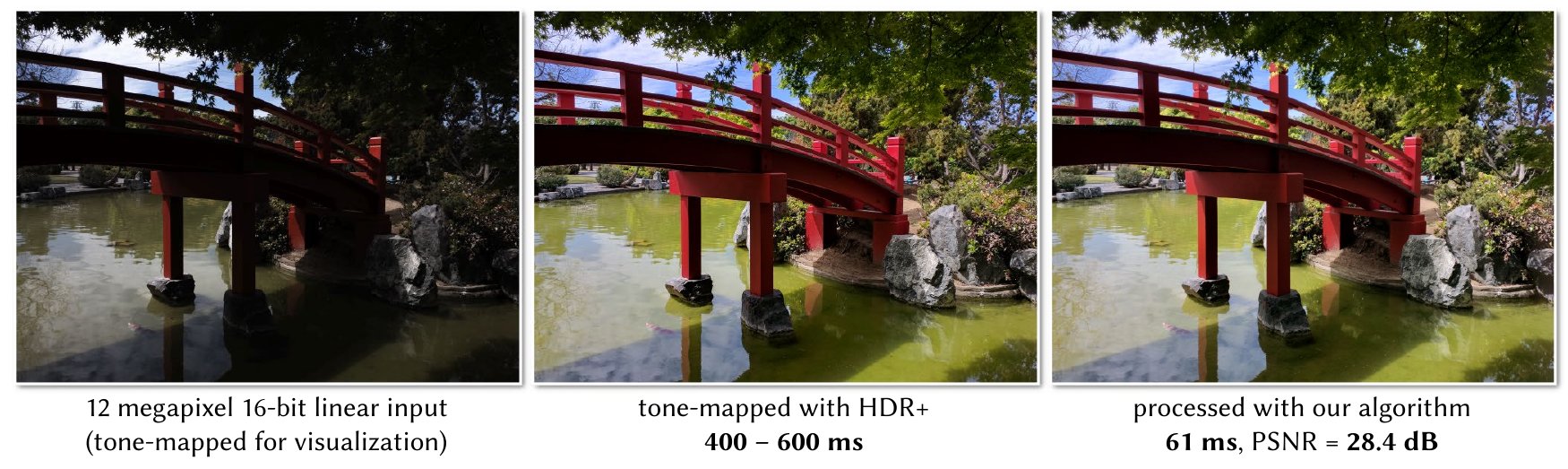}
  \caption{Our novel neural network architecture can reproduce sophisticated
  image enhancements with inference running in real time at full HD resolution
  on mobile devices. It can not only be used to dramatically accelerate reference
  implementations, but can also learn subjective effects from human retouching.}
  \label{fig:teaser}
\end{teaserfigure}

\begin{abstract}
Performance is a critical challenge in mobile image processing. Given a
reference imaging pipeline, or even human-adjusted pairs of images, we seek to
reproduce the enhancements and enable real-time evaluation.
For this, we introduce a new neural network architecture inspired by bilateral
grid processing and local affine color transforms. Using pairs of input/output
images, we train a convolutional neural network to predict the coefficients of
a locally-affine model in bilateral space.
Our architecture learns to make local, global, and content-dependent decisions to
approximate the desired image transformation.
At runtime, the neural network consumes a low-resolution version of the input
image, produces a set of affine transformations in bilateral space, upsamples
those transformations in an edge-preserving fashion using a new \emph{slicing}
node, and then applies those upsampled transformations to the full-resolution
image.
Our algorithm processes high-resolution images on a smartphone in
milliseconds, provides a real-time viewfinder at 1080p resolution, and matches
the quality of state-of-the-art approximation techniques on a large class of
image operators.
Unlike previous work, our model is trained off-line from data and therefore does
not require access to the original operator at runtime.
This allows our model to learn complex, scene-dependent transformations for
which no reference implementation is available, such as the photographic edits
of a human retoucher.

\end{abstract}

\setcopyright{acmlicensed}
\acmJournal{TOG}
\acmYear{2017}\acmVolume{36}\acmNumber{4}\acmArticle{118}\acmMonth{7}
\acmDOI{http://dx.doi.org/10.1145/3072959.3073592}

\ccsdesc[500]{Computing methodologies~Computational photography}
\ccsdesc[500]{Computing methodologies~Image processing}

\maketitle

\section{Introduction}

The high resolution of images and videos produced by contemporary cameras and
mobile devices puts significant performance pressure on image processing
algorithms, requiring sophisticated code optimization by skilled programmers.
While systems contributions have sought to facilitate the implementation of
high-performance executables,
\eg~\cite{JRK2013_halide,hegarty2014_darkroom,mullapudi2016_automatically},
they require programmer expertise, their runtime cost still grows with the
complexity of the pipeline, and they are only applicable when source code is
available for the filters.
Additionally, because image enhancement is subjective, it is often
desirable to learn an enhancement model directly from human adjustments,
\eg~\cite{Bychkovsky2011_mit5k}.
To this end, we present a machine learning approach where the effect of a
reference filter, pipeline, or even subjective manual photo adjustment is
learned by a deep network that can be evaluated quickly and with cost
independent of the reference's complexity. We focus on photographic enhancements that do not spatially warp the image or
add new edges, \eg~\cite{aubry2014_fast,hasinoff2016_burst}.

We share the motivation of prior work that seeks to accelerate ``black box''
image processing operations, either by using a remote server, \eg
\cite{Gharbi2015_recipe} or by processing a low-resolution image and then using
the low-resolution output to approximate a high-resolution
equivalent~\cite{Chen2016_bgu}.
For some operations, these approaches can achieve large speedups
%
but they suffer from significant limitations: the underlying image processing
operation must be somewhat scale-invariant
(Figure~\ref{fig:non_scale_invariant}), and must be fast to evaluate at low
resolution.
In addition, these techniques rely on the availability of an explicit reference
implementation, and therefore cannot be used to learn an implicitly-defined operation from a database of human annotated input/output pairs.

Many deep learning architectures have been used for image-to-image
transformations, \eg
\cite{long2015_fully,Xu2015_deep_edge_aware,liu2016_recursive,Yan2016_automatic,isola2016}.
However, most prior work incur a heavy computational cost that scales linearly
with the size of the input image, usually because of the large number of stacked
convolutions and non-linearities that must be evaluated at full resolution.
This general form allows for flexible models to be learned, but this
expressivity comes at a price: such architectures are orders of magnitude too
slow for real-time viewfinder applications, requiring seconds to process a $1$
megapixel image on the best desktop GPUs---more than $1000\times$ slower than
our proposed model ($2$\ms on GPU).
Our speedup is enabled by specifically targeting photographic transformations,
which are often well-approximated with linear operations in bilateral
space~\cite{Chen2016_bgu}, and accordingly learning our model in
this space.



We present a new network architecture that is capable of learning a
rich variety of photographic image enhancements and can be rapidly evaluated
on high-resolution inputs.
We achieve this through three key strategies:
\begin{inparaenum}[1)]
\item We perform most predictions in a \ADD{low-resolution} bilateral grid \cite{chen2007_real},
where each pixel's $x, y$ coordinates are augmented with a third dimension
which is a function of the pixel's color.
\ADD{To do this,} we introduce a new node for deep learning that performs a
data-dependent lookup.
This enables the so-called slicing operation, which reconstructs an output image
at full image resolution from the 3D bilateral grid by considering each pixel's
input color in addition to its $x, y$ location.
\item We follow previous work which has observed that it is often simpler to
predict the {\em transformation} from input to output rather than predicting the
output directly \eg, \cite{shih2013_data,Gharbi2015_recipe,Chen2016_bgu}.
This is why our architecture is designed to learn, as an intermediate
representation, a local affine color transformation that will be applied to the
input through a new multiplicative node.
\item While most of our learning and inference is performed at low resolution,
the loss function used during training is evaluated at full resolution, which
causes the low-resolution transformations we learn to be directly optimized
for their impact on high-resolution images.
\end{inparaenum}

Taken together, these three strategies (slicing, affine color transform, and
full-resolution loss) allow us to perform the bulk of our processing at a low
resolution (thereby saving substantial compute cost) yet reproduce the
high-frequency behavior of the reference operator.

We demonstrate the expressiveness of our model on a benchmark of 7
applications including: approximating published image filters
\cite{aubry2014_fast,hasinoff2016_burst}, reverse-engineering black-box
Photoshop actions, and learning the retouching style of
photographers~\cite{Bychkovsky2011_mit5k} from a set of manually corrected
photographs.
Our technique produces output whose quality is comparable to or better than
previous work, while being more widely applicable by not requiring some
reference implementation of the image operation being approximated, being
end-to-end learnable from input/output image pairs, and running in real-time on
mobile hardware.
The forward pass of our network takes $14$\ms to process a full screen resolution $1920\times1080$
image on a Google Pixel phone, thereby enabling real-time viewfinder effects at
$50$\hz.
\RM{We will release our code and trained models as open source.}

\section{Related Work}

Though image enhancement algorithms have been the focus of a great deal of
research, most sophisticated algorithms are too expensive to be evaluated
quickly on mobile devices, which is where the vast majority of digital images
are captured and processed.
%
Because of this, previous work has identified specific critical operations
and developed novel algorithms to accelerate them. For instance,
Farbman~et~al.~\shortcite{farbman2011_convolution} introduced
\emph{convolution pyramids} to accelerate linear translation-invariant
filters. Similarly, many approaches have been proposed to accelerate
bilateral filtering, due to the ubiquity of edge-aware image
processing~
\cite{tomasi1998_bilateral,paris2006_fast,chen2007_real,adams2010_fast}.

One way to accelerate an operator is to simply apply it at low resolution
and upsample the result.
A na{\"\i}ve upsampling will generally lead to an unacceptably blurry output, but
this issue can often be ameliorated by using a more sophisticated upsampling
technique that respects the edges of the original image.
Joint bilateral upsampling \cite{Kopf2007_joint_bilateral} does this by using a
bilateral filter on a high-resolution guidance map to produce a
piecewise-smooth edge-aware upsampling.
Bilateral space optimization \cite{barron2015_fast,barron2015_solver} builds
upon this idea by solving a compact optimization problem inside a bilateral
grid, producing upsampled results which are maximally smooth.

Gharbi~et~al.~\shortcite{Gharbi2015_recipe} focus on learning the
\emph{transformation} from input to output instead of the output itself. They
approximate a large class of complex, spatially-varying operators with a
collection of simple local models---a \emph{transform recipe}---that is
tailored to a given input/output pair. The task of computing the operator and
fitting the recipe is offloaded to the cloud while the mobile device need
only apply the recipe, thereby saving time and energy.
Similarly, Chen~et~al.~\shortcite{Chen2016_bgu} approximate an image operator
with a grid of local affine models in bilateral space, the parameters
of which are fit to an input/output pair in a manner resembling the guided
filter~\cite{He2013_guided}. By performing this model-fitting on a
low-resolution image pair, this technique enables real-time on-device
computation.
We build upon this bilateral space representation, but rather than fitting a
model to approximate a single instance of an operator from a pair of
images, we construct a rich CNN-like model that is trained to apply the
operator to any unseen input.
This bypasses the need for the original operator at runtime and opens up the
opportunity to learn non-algorithmic transformations (\ie, hand-adjusted
input/output image pairs).
This also allows us to optimize the affine coefficients to model the operator
running at full resolution, which is important for filters that vary with scale
(Figure~\ref{fig:non_scale_invariant}).

\paragraph{Neural networks for image processing}
Recently, deep convolutional networks have achieved significant progress on
low-level vision and image processing tasks such as depth estimation
\cite{eigen2014_depth}, optical flow \cite{ilg2016_flownet},
super-resolution \cite{dong2014_superres}, demosaicking and denoising
\cite{Gharbi2016_demosaicnet,zhang2016_noise}, image matting
\cite{shen2016_portrait}, colorization \cite{iizuka2016_color}, and
general image-to-image ``translation'' tasks \cite{isola2016}.
Recent work has even explored learning deep networks within a bilateral grid
\cite{Jampani2016} though this work does not address our task of learning
image transformations in that space, and instead focuses on classification and
semantic segmentation.
Some architectures have been trained to approximate a general class of
operators. Xu~et~al.~\shortcite{Xu2015_deep_edge_aware} develop a three-layer
network in the gradient domain to accelerate edge-aware smoothing filters.
Liu~et~al.~\shortcite{liu2016_recursive} propose an architecture to learn
recursive filters for denoising, image-smoothing, inpainting and color
interpolation. They jointly train a collection of recursive networks and a
convolutional network to predict image-dependent propagation weights.
While some of this work can process low-resolution images on a desktop GPU at
interactive rates, they remain too slow for our application: real-time
processing of high-resolution images on a mobile device.

\paragraph{Automatic photo editing}

Our model can be trained to automatically correct photographs from
input/output image pairs provided by a human retoucher.
This is the task introduced by Bychkovsky~et~al.
\shortcite{Bychkovsky2011_mit5k}, who estimate global brightness/contrast
adjustments that characterize the personal style of 5 trained photographers.
They train a regression model with handcrafted features that capture both
low-level information and semantic content (\eg, faces) on a dataset of
\num{5000} raw images. Hwang~et~al.~\shortcite{hwang2012_context} approach the
problem with a coarse-to-fine search for the best-matching scenes that takes
more than a minute for a $500\times333$ image.
Kaufman~et~al.~\shortcite{Kaufman2012_CAPE} learn local color and
contrast manipulations from hard-coded features (faces, blue skies, clouds,
underexposed areas), running over 2 minutes for a VGA image.
More recently, Yan~et~al.~\shortcite{Yan2016_automatic} use a compact pixel-wise
neural network and handcrafted features. Their network takes $1.5$\secs to
process a 1~megapixel image (on top of the time needed for object detection,
dense image segmentation, and scene recognition used in their features).
Our model can learn similar global tonal adjustments and generalizes to more
complex effects, including color corrections and local edits, in addition to
being much faster.

\begin{figure*}[!t]
    \centering
    \includegraphics[width=\textwidth]{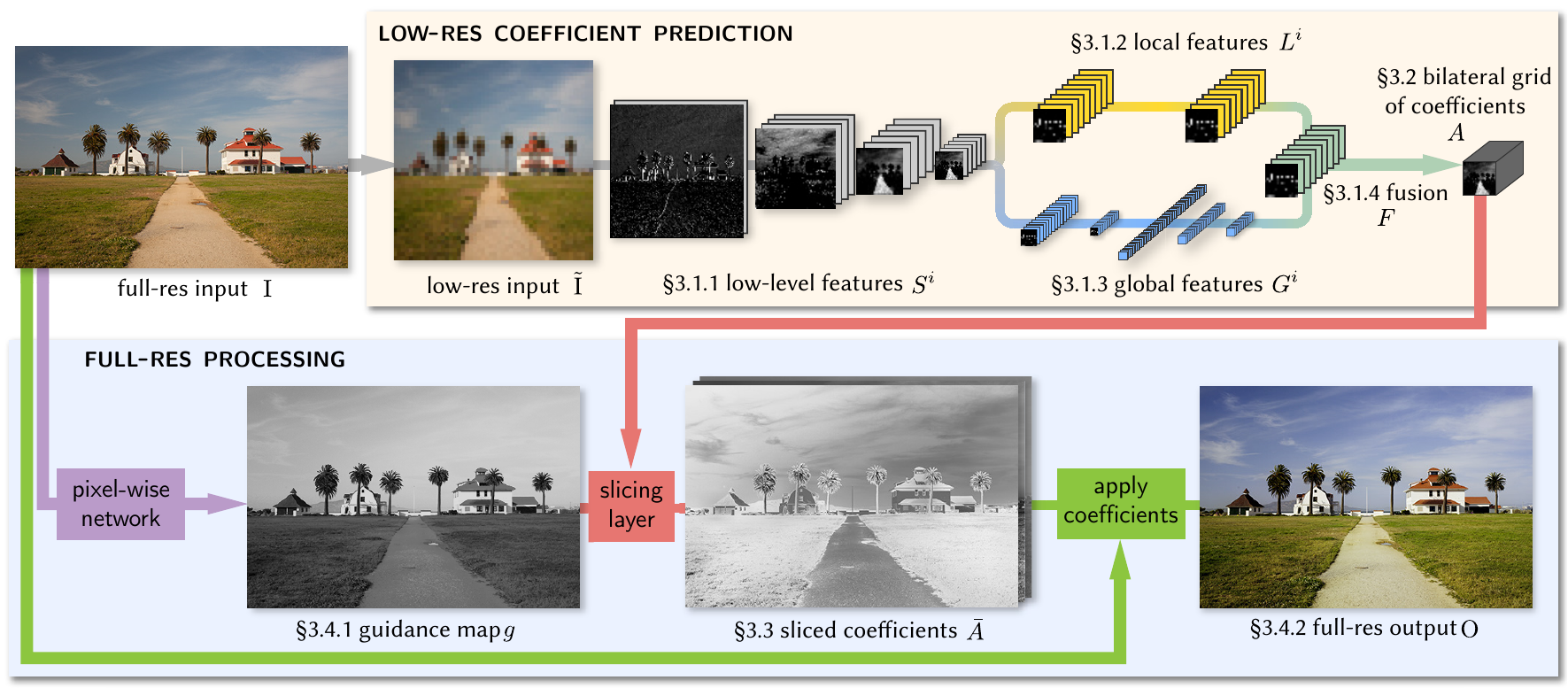}
    \caption{
  Our new network architecture seeks to perform as much computation as possible
  at a low resolution, while still capturing high-frequency effects at full
  image resolution. It consists of two distinct streams operating
      at different resolutions.
	  The \emph{low-resolution} stream (top) processes a
      downsampled version \DownsampledInput of the input \Input through several convolutional layers so
      as to estimate a bilateral grid of affine coefficients \ParameterLayer.
      This low-resolution stream is further split in two paths to learn both
      local features $\Local^i$ and global features $\Global^i$, which are fused
      (\Fused) before making the final prediction.
	  The global and local
      paths share a common set of low-level features $\Shared^i$.
      In turn, the \emph{high-resolution} stream (bottom) performs a minimal yet
      critical amount of work: it learns a grayscale guidance map \Guide used by our new
      \emph{slicing} node to upsample the grid of affine coefficients back to
      full-resolution \SlicedParameterLayer. These per-pixel local affine
      transformations are then applied to the full-resolution input, which yields the final output
      \Output.}
    \label{fig:network}
\end{figure*}

\section{Our architecture}
\label{sec:network}

We propose a new convolutional network architecture that can be trained to
perform fast image enhancement (Figure~\ref{fig:network}).
Our model is designed to be expressive, preserve edges, and require
limited computation at full resolution.
It is fully end-to-end trainable and runs in real-time at 1080p on a
modern smartphone.

We perform most of the inference on a low-resolution copy \DownsampledInput
of the input \Input in the {\em low-res} stream (Fig.~\ref{fig:network}, top),
which ultimately predicts local affine \RM{color} transforms in a representation
similar to the bilateral grid \cite{Chen2016_bgu}.
In our experience, image enhancements often depend not only on local image
features but also on global image characteristics such as histograms, average
intensity, or even scene category.
Therefore, our low-res stream is further split into a \ADD{{\em local} path} and a
\ADD{{\em global} path}. Our architecture then fuses these two paths to yield the final
coefficients representing the affine transforms.

The {\em high-res} stream (Fig.~\ref{fig:network}, bottom) works at full
resolution and performs minimal computation
but has the critical role of capturing
high-frequency effects and preserving edges when needed. For this purpose, we introduce a
slicing node inspired by bilateral grid processing
\cite{paris2006_fast,chen2007_real}. This node performs data-dependent lookups
in the low-resolution grid of affine coefficients based on a learned
\ADD{\emph{guidance map}}.
Given high-resolution affine coefficients obtained by slicing into the grid with
the full-resolution guidance map, we apply local color transforms to each pixel
to produce the final output \Output.
%
At training time, we minimize our loss function at \emph{full resolution}. This
means that the low-res stream, which only processes heavily downsampled data,
still learns intermediate features and affine coefficients that can reproduce
high-frequency effects.

As a first approximation, one can think of our work as alleviating the need for
the reference filter at runtime in Chen~et~al.'s Bilateral Guided Upsampling
\shortcite{Chen2016_bgu}.
In a sense, we seek to predict the affine color transform coefficients in the
bilateral grid given a low-resolution version of the image.
However, there are several key elements that go beyond this. First,
the downsampling into the bilateral grid is learned. Second, the guidance image
is also learned and not restricted to luminance. Finally, we apply the loss
function not on the affine coefficients, but on the final image at full
resolution, which allows us to capture high-frequency effects and handle
operators that are not scale-invariant (Figure~\ref{fig:non_scale_invariant}).
\ADD{We illustrate the role of each component of our architecture with an ablation
study in Figures~\ref{fig:with_without_splat}, \ref{fig:with_without_global},
\ref{fig:with_without_slicing} and \ref{fig:with_without_learned_guide}.}

\subsection{Low-resolution prediction of bilateral coefficients}
\label{sec:features_subnetwork}

\ADD{The input \DownsampledInput to the low-res stream has a fixed resolution $256\times 256$.}
\ADD{It is first processed by a stack of strided convolutional layers
$(\Shared^i)_{i=1,\ldots,\NShared}$ to extract \emph{low-level} features and reduce
the spatial resolution.}
Then, in a design inspired by Iizuka~et~al.~\shortcite{iizuka2016_color},
\ADD{the last low-level features are processed by} two asymmetric \ADD{paths}:
the first \ADD{path $(\Local^i)_{i=1,\ldots,\NLocal}$} is fully convolutional \cite{long2015_fully} and
specializes in learning local features that propagate image data while retaining
spatial information.
The second \ADD{path $(\Global^i)_{i=1,\ldots,\NGlobal}$} \RM{follows the design
of standard classification networks \cite{krizhevsky2012_imagenet} and learns}
uses \ADD{both convolutional and} fully-connected layers to learn a fixed-size
vector of global features (\eg high-level scene category, indoor/outdoor, etc.)
with a receptive field covering the entire low-resolution image \DownsampledInput.
The outputs of the two \ADD{paths}, $\Global^{\NGlobal}$ and $\Local^{\NLocal}$, are
then fused into a common set of features \Fused.
A pointwise linear layer outputs a final array \ParameterLayer from the fused
streams.
We interpret this array as a bilateral grid of affine coefficients
(Section~\ref{sec:features_as_bilateral}). Since we produce a 3D bilateral
grid from a 2D image in a content-dependent fashion, we can view the
\emph{low-res} stream as implementing a form of \emph{learned splatting}.
\RM{These layers progressively
reduce the spatial resolution through a series of strided
convolutions with stride $\Stride=2$; we double the number of features each
time. We use $\NShared=4$, for at total reduction of $2^{\NShared} = 16$ in each
spatial dimension.}
\RM{akin to Siamese networks
\cite{bromley1993_siamese}. In practice, these shared layers process
a unique input, the low-resolution image \DownsampledInput of size $256\times 256$,
instead of the usual two distinct inputs of Siamese networks.}
\RM{thus the last layer \ADD{before the fork} has a $16\times 16$ resolution.}
\RM{The last layer of \ADD{low-level} features $\Shared^{\NShared}$}
\RM{The last layer before the fork, $\Shared^{\NShared}$, is then separately processed by
\begin{inparaenum}[1.] \item $\NLocal=2$ convolutional layers $\Local^i$ in the local
  \ADD{path}, \item and by
$\NGlobal=5$ layers $\Global^i$ in the global \ADD{path}, of which the first two
are strided convolutions, and the remaining 3 are fully connected.
\end{inparaenum}}

\subsubsection{\ADD{Low-level features}}
We first process the low-resolution image $\Shared^0 := \DownsampledInput$ with
a stack of standard strided convolutional layers with stride
$\Stride=2$ \ADD{(Figure~\ref{fig:network})}:
\RM{Each layer $\Shared^i$ is computed from the previous one as follows:}
\begin{equation}
\resizebox{3.05in}{!}{$ \displaystyle
  \Shared^i_c[x,y] = \Activation\left(
  \Bias^i_{c}+
  \sum_{\substack{x',y',c'}}
  \Weight^i_{cc'}\!\left[ x', y' \right]\Shared^{i-1}_{c'}\left[\Stride x+x',\Stride y+y' \right]\right)
  \label{eq:strided_convolution}
$}
\end{equation}
Where $i=1,\ldots,\NShared$ indexes the layers, $c$ and $c'$ index the
layers' channels, $\Weight^i$ is an array of weights for the convolutions, $\Bias^i$ is a
vector of biases, and the summation is over $-1 \leq x',y' \leq 1$ (\ie, the
convolution kernels have $3\times3$ spatial extent).
We use the ReLU activation function $\Activation(\cdot)=\max(\cdot, 0)$ and
use zero-padding as the boundary condition in all convolutions.

\begin{figure}[!b]
    \centering
    \includegraphics[width=.97\columnwidth]{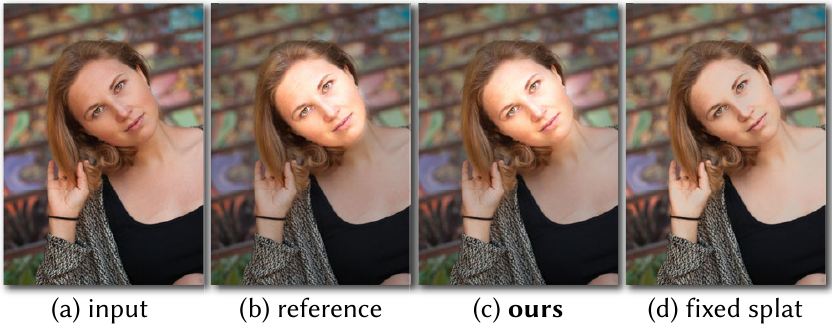}
    \caption{
      Our \emph{low-level} convolutional layers are fully learned and can
      extract semantic information. Replacing these layers with the standard
      bilateral grid splatting operation causes the network to lose much of its
      expressive power. In this example of our \emph{Face brightening} operator (a-b),
      the network with hardcoded splatting (d) cannot detect the face properly
      because the grid's resolution is too low. Instead, it
      slightly brightens all skintones, as is visible on the hands. Our
      progressive downsampling with strided convolutions learns the
      semantic features required to solve this task properly (c), brightening
      only the face while darkening the background like in the reference.}
    \label{fig:with_without_splat}
\end{figure}

\ADD{These low-level layers progressively reduce the spatial dimensions by a
total factor of $2^{\NShared}$. Thus \NShared has two effects:
\begin{inparaenum}[1)]
\item it drives the spatial downsampling between the
low-resolution input \DownsampledInput and the final grid of affine coefficients---the
higher \NShared, the coarser the final grid, and
\item \NShared controls the complexity of the prediction:
deeper layers have an exponentially larger spatial support and more
complex non-linearities (by composition); thus, they can extract more complex
patterns in the input.
\end{inparaenum}}
Figure~\ref{fig:with_without_splat} shows a comparison with a network in which the
low-level layers have been removed, and replaced by a hard-coded splatting
operation~\cite{chen2007_real}. Without \ADD{these layers}, the
network loses much of its expressive power.
Our architecture, uses $\NShared=4$ low-level layers.
\ADD{Table~\ref{tab:architecture} summarizes the dimensions of each layer.}
\RM{$\Shared^{1}$ has 8
feature channels, and we double the number of features
with each progressive reduction in spatial dimension. This leads to a final
low-level layer $\Shared^{\NShared}$ with dimensions $16\times 16\times 64$.}
\RM{By reducing the spatial dimensions of the low-res input, the shared layers
effectively learn a mapping from 2D image to 3D bilateral space
(Section~\ref{sec:features_as_bilateral})}

\subsubsection{Local \ADD{features path}}
The last low-level features layer $\Shared^{\NShared}$ is then processed by a stack of
$\NLocal=2$ convolutional layers $\Local^i$ in the local \ADD{path}
\ADD{(Figure~\ref{fig:network}, yellow)}. These layers take the same form
as Equation~\eqref{eq:strided_convolution}, identifying $\Local^0 :=
\Shared^{\NShared}$, but this time with stride $\Stride=1$.
We keep both the spatial resolution and number of features constant in the local
\ADD{path}. Because the resolution is held constant, the spatial support of the
filters only grows linearly with \NLocal.
A deep enough stack of convolution layers, roughly measured by
$\NShared+\NLocal$, is critical to capturing useful semantic
features~\cite{krizhevsky2012_imagenet}. If a higher spatial resolution is
desired for the final grid of coefficients, one can reduce \NShared and
increase \NLocal to compensate accordingly, so as not to reduce the
expressiveness of the network.
Without the local \ADD{path}, the predicted coefficients would \ADD{lose any
notion} of spatial location.

\subsubsection{Global \ADD{features path}}
\ADD{Like the local path, the global features path} branches out from
$\Shared^{\NShared}$, that is $\Global^0 := \Shared^{\NShared}$. 
It comprises two strided convolutional layers
(Equation~\eqref{eq:strided_convolution}, with $\Stride=2$)
followed by three fully-connected layers, for a total of $\NGlobal=5$ global
layers \ADD{(Figure~\ref{fig:network}, blue)}.
\RM{$\Global^{2}$ has dimensions $4\times 4\times 64$, which we flatten into a
1024-dimensional vector.\RM{(keeping the same notation for both forms)}
$\Global^{3}$, $\Global^{4}$, $\Global^{5}$ have dimensions 256, 128 and 64
respectively.}
One consequence of using fully-connected layers is that the resolution of the
input \DownsampledInput needs to be fixed, \ADD{since it dictates the dimensions of
$\Global^2$ and the number of network parameters that act on it}.
As we will see in Section~\ref{sec:trainable_slicing}, thanks to our slicing
operator, we can still process images of \ADD{any resolution}, despite the size
of the \emph{low-res} stream being fixed. 
\RM{The strided convolutions  $\Global^1$ and $\Global^2$ further reduce the spatial
dimensions, \ADD{but this time} the number of features \ADD{is kept} constant.
These layers take the same form as Equation~\eqref{eq:strided_convolution},
with $\Global^0 = \Shared^{\NShared}$ and stride $\Stride=2$.}
\RM{
$\Global^{2}$ has dimensions $4\times 4\times 64$, which we flatten into a
1024-dimensional vector (keeping the same notation for both forms).
The next three layers are fully connected, with 256, 128 and 64 hidden nodes.
respectively:
\begin{align}
  \Global^i_c = \Activation\left(\Bias_c + \sum_{c'}
  \Weight_{cc'}\Global^{i-1}_{c'}\right) & &\text{for}\ i = 3,\ldots,\NGlobal\
  \label{eq:fully_connected}
\end{align}}

\begin{figure}[!t]
    \centering
    \includegraphics[width=\columnwidth]{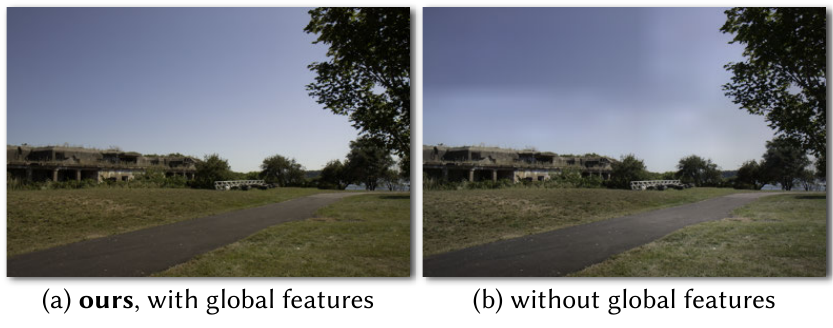}
    \caption{
    The \ADD{global features path} in our architecture allows our model to reason
    about the full image, \eg, for subjective tasks such as reproducing
    subjective human adjustments that may be informed by intensity distribution
    or scene type (a). Without the \ADD{global path, the model can make local
    decisions that are spatially inconsistent (b). Here, the network fails to
    recognize that the blue area in the top-left corner also belongs to the sky and
    should therefore receive the same correction as the area just below it.}
	}
    \label{fig:with_without_global}
\end{figure}

The global \ADD{path} produces a 64-dimensional vector that summarizes global
information about the input and acts as a prior to regularize the local decisions
made by the local \ADD{path}. Without global features \ADD{to encode this
high-level description} of the input, the network can make erroneous local
decisions that lead to artifacts \ADD{as exemplified by the large-scale variations
in the sky in Figure~\ref{fig:with_without_global}}.
 
\subsubsection{Fusion and linear prediction}
We fuse the contributions of the local and global \ADD{paths} \ADD{with} a
pointwise affine mixing followed by a ReLU activation:
\begin{equation}
  \Fused_c[x,y] = \Activation\left( \Bias_{c} +
  \sum_{c'} {\Weight'}_{cc'}\Global^{\NGlobal}_{c'}+
  \sum_{c'} \Weight_{cc'}\Local^{\NLocal}_{c'}[x,y]\right) 
  \label{eq:fusion}
\end{equation}
\RM{This operation can be interpreted as \ADD{concatenating} a copy of the global feature
vector $\Global^{\NGlobal}$ at each pixel of the local feature map
$\Local^{\NLocal}$ then performing a $1\times 1$ convolution followed by a
rectifier.}
This yields a $16\times 16\times 64$ array of features from \ADD{which}, we
make our final $1\times 1$ linear prediction to produce a $16\times 16$ map with
96 channels:
\begin{equation}
  \ParameterLayer_c[x,y] = \Bias_{c} + \sum_{c'} \Fused_{c'}[x,y]\Weight_{cc'}
  \label{eq:parameters_linear_prediction}
\end{equation}

\begin{table}[!htp]
\caption{
\ADD{
  Details of the network architecture. \textit{c}, \textit{fc}, \textit{f} and
  \textit{l} refer to convolutional, fully-connected, fusion and pointwise
  linear layers respectively.}
}
\label{tab:architecture}
\centering
\tabcolsep=0.11cm
\begin{tabular}{l*{4}{c}|*{2}{c}|*{5}{c}|*{2}{c}}
         & $\Shared^1$ & $\Shared^2$ & $\Shared^3$ & $\Shared^4$ & $\Local^1$ & $\Local^2$ & $\Global^1$ & $\Global^2$ & $\Global^3$ & $\Global^4$ & $\Global^5$ & $\Fused$   & $\ParameterLayer$ \\
  \hline
  type   & \textit{c}  & \textit{c}  & \textit{c}  & \textit{c}  & \textit{c} & \textit{c} & \textit{c}  & \textit{c}  & \textit{fc} & \textit{fc} & \textit{fc} & \textit{f} & \textit{l} \\
size    & 128         & 64          & 32          & 16          & 16         & 16         & 8           & 4           & \na         & \na         & \na         & 16         & 16\\
channels & 8           & 16          & 32          & 64          & 64         & 64         & 64          & 64          & 256         & 128         & 64          & 64         & 96\\
\end{tabular}
\end{table}

\subsection{Image features as a bilateral grid}
\label{sec:features_as_bilateral}

So far we have described our model as a neural network. We now shift our
perspective to that of a bilateral grid.
To facilitate this, in a slight abuse of notation, we will occasionally treat
the final feature map \ParameterLayer as a multi-channel bilateral grid whose
third dimension has been unrolled:
\begin{equation}
\ParameterLayer_{\GridDepth c+z}[x,y] \leftrightarrow \ParameterLayer_c[x,y,z]
  \label{eq:unfold_grid}
\end{equation}
where $\GridDepth=8$ is the depth of the grid.
Under this interpretation, \ParameterLayer can be viewed as a
$16\times 16\times 8$ bilateral grid, where each grid cell contains $12$ numbers,
one for each coefficient of a $3\times4$ affine color transformation matrix.
This reshaping \RM{operation} lets us interpret the strided convolutions in
Equation~\eqref{eq:strided_convolution} as acting in the bilateral domain, where
they correspond to a convolution in the $(x,y)$ dimensions and express full
connectivity in the $z$ and $c$ dimensions.
This operation is therefore more expressive than simply applying 3D
convolutions in the grid, which would only induce local connectivity on
$z$~\cite{Jampani2016}.
It is also more expressive than standard bilateral grid splatting \ADD{which
discretizes \Input into several intensity bins then box filters the result
\cite{chen2007_real}; an operation that is easily expressed with a 2-layer
network.}
In a sense, by maintaining a 2D convolution formulation throughout and only
interpreting the last layer as a bilateral grid, we let the network decide when
the 2D to 3D transition is optimal.

\subsection{Upsampling with a trainable slicing layer}
\label{sec:trainable_slicing}

So far we have described how we learn to predict a bilateral grid of
coefficients \ParameterLayer from a low-resolution image \DownsampledInput using
the \emph{low-res} stream of our network.
We now need to transfer this information back to the high-resolution space of
the original input \Input to produce our final output image.
To this end, \ADD{we introduce a layer based on the bilateral
grid \emph{slicing} operation \cite{chen2007_real}. 
This layer takes as input a single-channel guidance map \Guide and a
feature map \ParameterLayer (viewed as a bilateral grid)
with a much lower spatial resolution than \Guide.
It performs a data-dependent lookup in the final feature map
\ParameterLayer.
The layer is sub-differentiable with respect to both \ParameterLayer and
\Guide. This allows us to backpropagate through it at train time.}

The result of the slicing operator is a new feature map \SlicedParameterLayer
with the same spatial resolution as \Guide, obtained by tri-linearly
interpolating the coefficients of \ParameterLayer at locations defined by
\Guide:
\begin{equation}
  \SlicedParameterLayer_c[x,y] =
  \sum_{i,j,k}\tent\left(\WidthRatio x - i\right)\tent\left(\HeightRatio y -
  j\right)\tent \left( \GridDepth\cdot\Guide[x,y]-k \right)
  \ParameterLayer_c\left[i, j, k\right]
  \label{eq:slicing}
\end{equation}
Using a linear \ADD{interpolation kernel} $\tent(\cdot) = \max(1-|\cdot|, 0)$,
and where \WidthRatio and \HeightRatio are the width and height ratios of the
grid's dimensions w.r.t.\ the full-resolution image's dimensions. 
\ADD{Essentially, each pixel is assigned the vector of coefficients whose
depth in the grid is given by the gray scale value $\Guide[x,y]$, \ie,
loosely speaking $\ParameterLayer_c[i, j, \Guide[x,y]]$.}
\ADD{Flownet2~\cite{ilg2016_flownet} and Spatial Transformer
Networks~\cite{jaderberg2015_spatial} have used similar interpolation operators
for in-network spatial warping.}
We fix the spatial resolution of the grid to $16\times 16$, and its depth to
$\GridDepth=8$.

\begin{figure}[!t]
    \centering
    \includegraphics[width=\columnwidth]{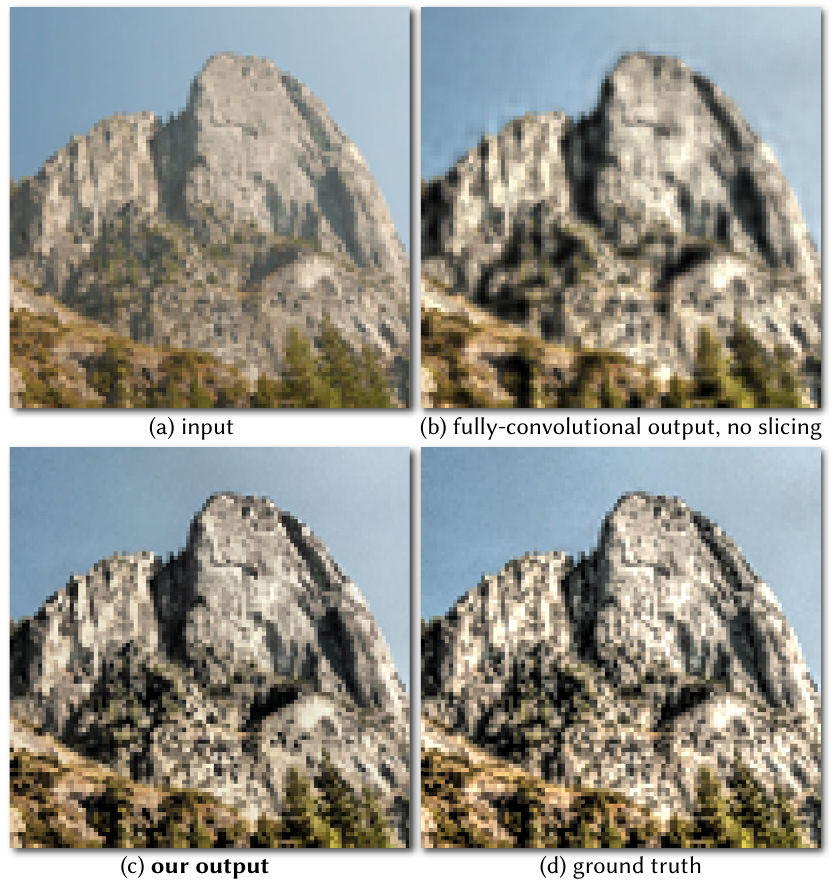}
   \caption{
    Our new slicing node is central to the expressiveness of our architecture and its handling of high-resolution effects. 
      Replacing this node with a standard bank of learnable deconvolution filters
      reduces expressiveness (b) because no full-resolution data is
      used to predict the output pixels.
      Thanks to its learned full-resolution guidance map, our slicing layer 
      approximates the desired enhancement with much higher fidelity (c), thereby preserving
      the edges of the input (a) and capturing the high-frequency
      transformations visible in the ground-truth output (d).
      }
    \label{fig:with_without_slicing}
\end{figure}

The slicing operation is parameter-free and can be implemented efficiently in an
OpenGL shader~\cite{chen2007_real}.
It acts as a bottleneck layer that constrains the representation of the neural
network to a low-dimensional space.
This both simplifies the learning problem and speeds up the processing
time~\cite{barron2015_fast,barron2015_solver}.
Crucially, performing inference within a bilateral grid forces our model's
predictions to follow the edges in \Guide, thereby regularizing our
predictions towards \emph{edge-aware} solutions \ADD{(unlike standard networks
based on transpose-convolutions or ``deconvolution layers'',
Figure~\ref{fig:with_without_slicing})}.
This design decision tends to benefit photographic manipulation tasks such as
ours and enables our significant speedup over more general models \ADD{due to the
low dimensionality of \ParameterLayer (Figure~\ref{fig:performance_comparison}).}

This data-dependent lookup is critical to the expressive power of our model.
As we will see in Section~\ref{sec:assembling_output}, it allows us to predict
a complex operation on the full-resolution image using a collection of much
simpler local models.

\begin{figure}[!b]
    \centering
	\begin{tabular}{@{}lr@{}}
	    \includegraphics[width=.48\columnwidth]{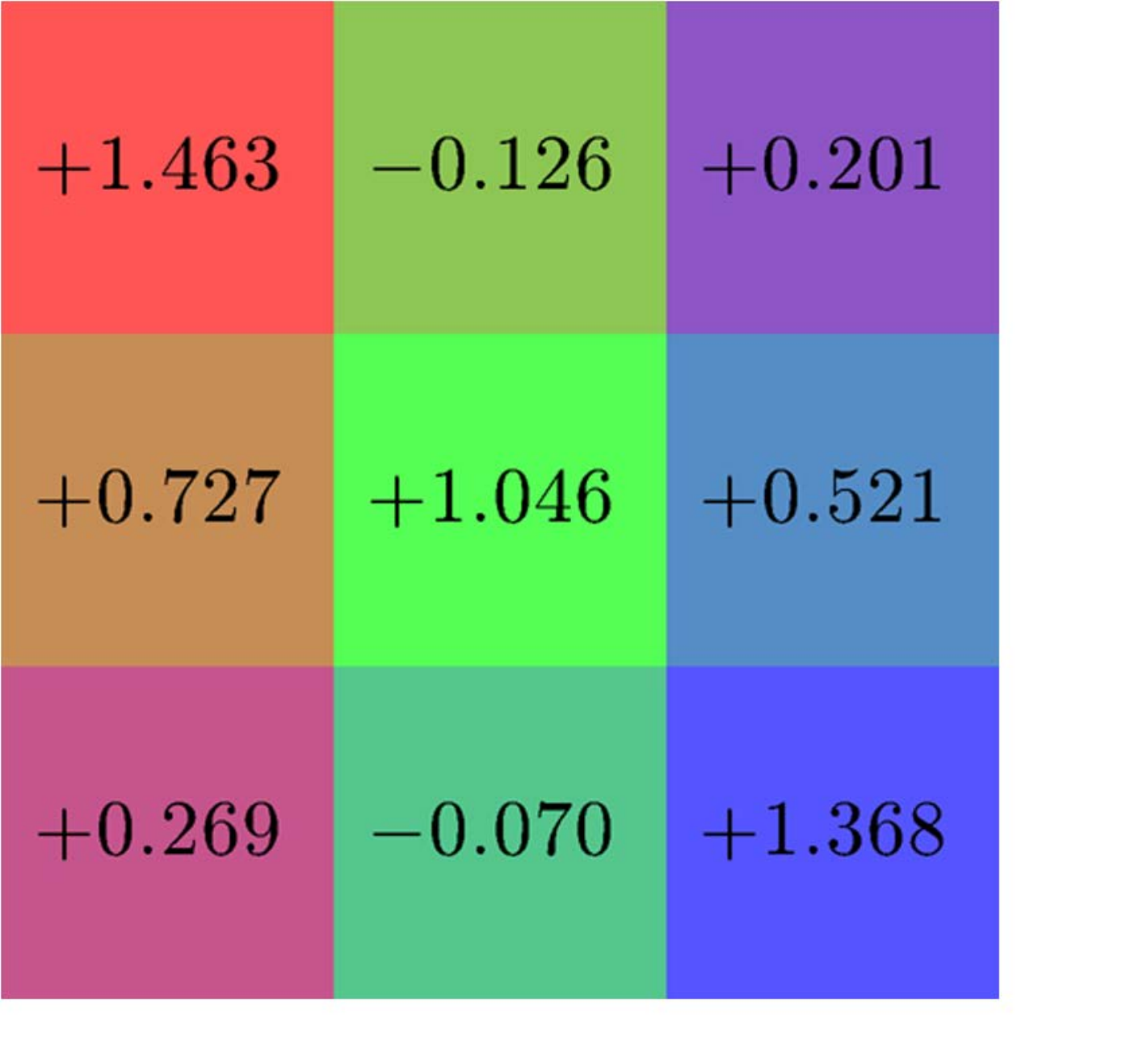}
	    \includegraphics[width=.48\columnwidth]{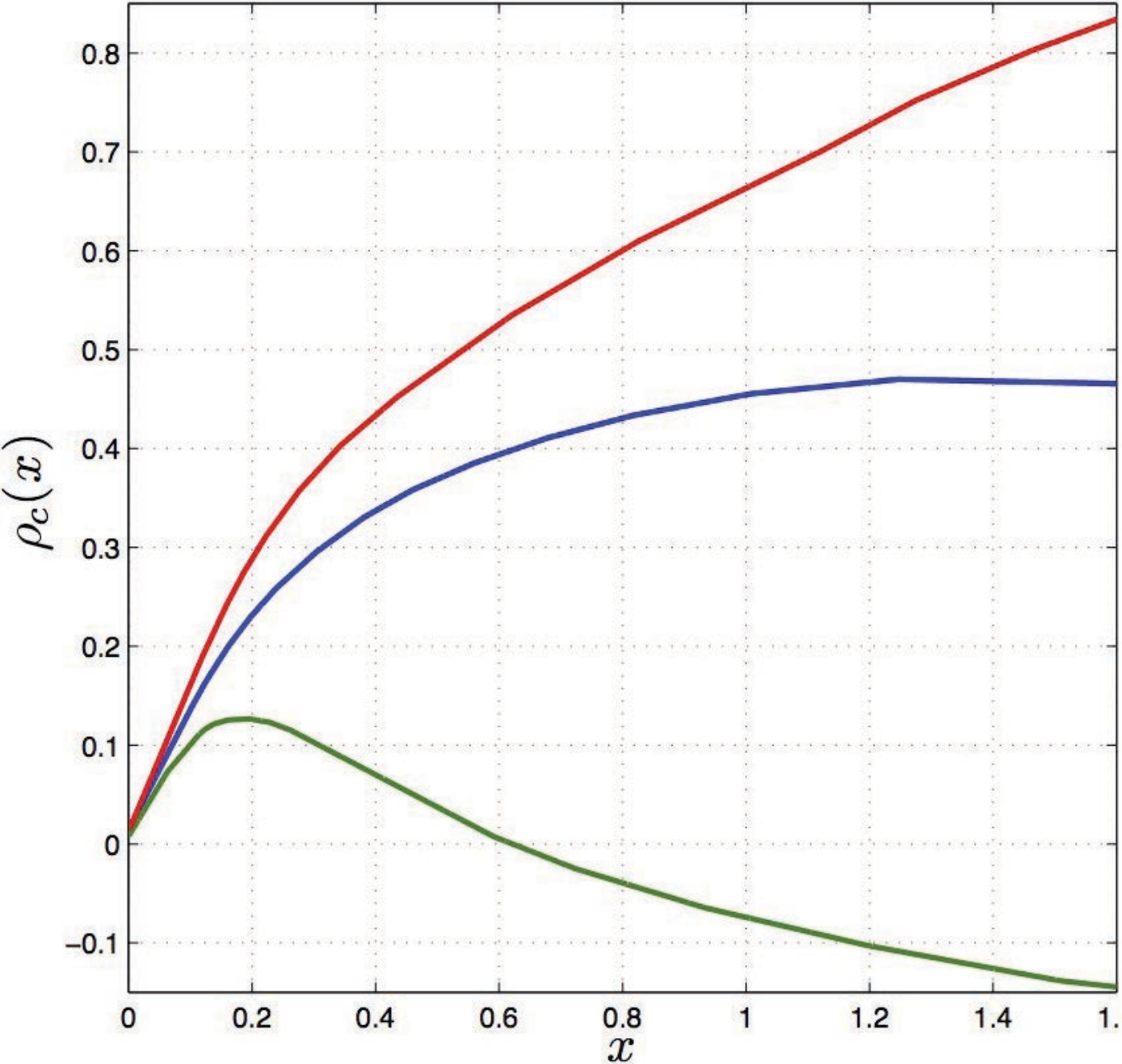} &
	\end{tabular}
    \caption{The color transform matrix (left) and per-channel tone curves
      (right) used to produce the guidance map \Guide, as learned by one
      instance of our model.
	}
    \label{fig:tonecurve_and_ccm}
\end{figure}

\subsection{Assembling the full-resolution output}
\label{sec:fullres}

\ADD{So far, we have described how to obtain and upsample the bilateral grid of
affine coefficients. The rest of the processing is done at full resolution.}
It \ADD{should} therefore be simple and easily-parallelizable to minimize
computational cost.
\ADD{From the full-resolution input \Input, we extract a set of
$\NFullresFeatures$ full-resolution features \FullresFeatures that fulfill two roles:
\begin{inparaenum}[1)]
\item they are combined to predict the guidance map \Guide used in the slicing node, and
\item they are used as regression variables for the local affine models.
\end{inparaenum}
}

\ADD{The most cost-efficient approach is to use the channels of the 
input image as features, that is $\FullresFeatures=\Input$ (with
$\NFullresFeatures=3$) and the local affine models are color
transformations. All our results use this fast formulation.}

\subsubsection{Guidance map auxiliary network}
\label{sec:guide_subnetwork}

\begin{figure}[!t]
    \centering
    \includegraphics[width=\columnwidth]{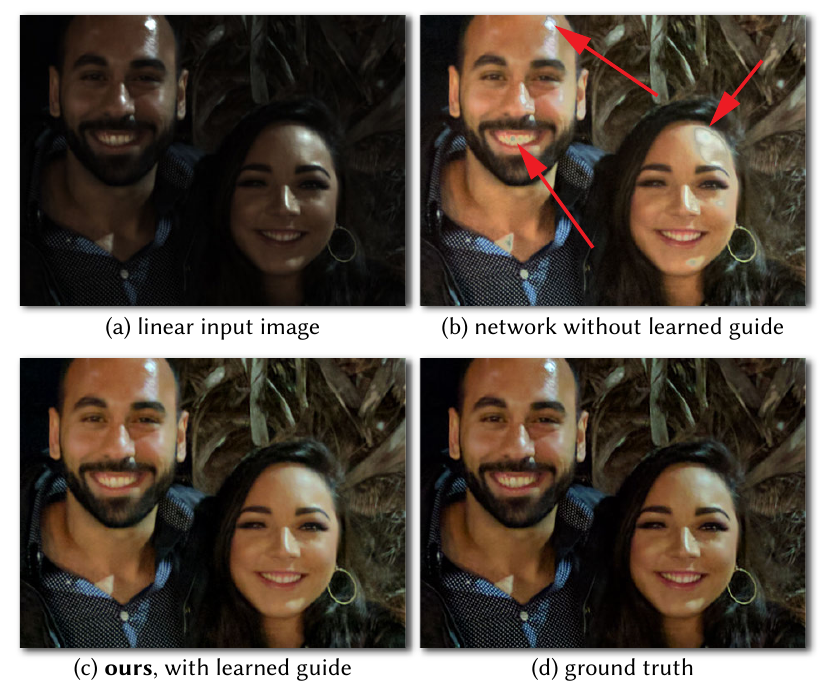}
    \caption{
	Our slicing node uses a learned guidance map.
	Using luminance as guide causes artifacts with the HDR+ pipeline
	reproduction, in particular with posterization artifacts in the highlights on
	the forehead and cheeks (b).
	In contrast, our learned guide (c) correctly reproduces the ground truth (d).
	}
    \label{fig:with_without_learned_guide}
\end{figure}

We define \Guide as a simple pointwise \ADD{nonlinear} transformation \ADD{of
the full-resolution features}:
\begin{equation}
  \Guide[x,y] = \Bias + \sum_{c=0}^{2}
  \Curve_c \left(\ColorCorrectionMatrix_c^\top\cdot\FullresFeatures_c[x,y] + \Bias_c' \right)
  \label{eq:guide}
\end{equation}
Where $\ColorCorrectionMatrix^\top_c$ are the rows of a $\InputChannels\times
\InputChannels$ color transformation matrix, \Bias and $\Bias'_c$ are scalar biases,
and $\Curve_c$ are piecewise linear transfer functions parametrized as a
\ADD{sum} of $\NCurvePoints$ scaled ReLU functions with thresholds
$\CurveThreshold_{c,i}$ and slopes $\CurveSlope_{c,i}$:
\begin{equation}
  \Curve_c(x) = \sum_{i = 0}^{15} \CurveSlope_{c,i}\max \left( x-\CurveThreshold_{c,i}, 0 \right)
  \label{eq:piecewise_linear_curve}
\end{equation}
The parameters \ColorCorrectionMatrix, \CurveSlope, \CurveThreshold, \Bias,
$\Bias'$ are learned jointly with the other network parameters.
\ColorCorrectionMatrix is initialized to the identity and \CurveSlope,
\CurveThreshold, \Bias, and \Bias' are initialized such each $\Curve_c$ is an
identity mapping over $[0,1]$, which is necessary to avoid learning a
degenerate \Guide.
Figure~\ref{fig:with_without_learned_guide} shows the impact of using this
learned guide and Figure~\ref{fig:tonecurve_and_ccm} shows an example of the
color transformation matrix and tone curve that are learned for \ADD{the
corresponding} task.




\subsubsection{Assembling the final output}
\label{sec:assembling_output}

Although image operators may be complex when viewed at the scale of an entire
image, recent work has observed that even complicated image processing pipelines
can often be accurately modeled as a collection of simple local
transformations~\cite{He2015_fast_guided,Gharbi2015_recipe,Chen2016_bgu}.
We therefore model each channel of our final output $\Output_c$ as an affine
combination of the \ADD{full-resolution features}, with coefficients defined
by the channels of the sliced feature map \SlicedParameterLayer:
\ADD{
\begin{equation} \Output_c[x,y] = \SlicedParameterLayer_{\NFullresFeatures + (\NFullresFeatures+1)c}
  + \sum_{c'=0}^{\NFullresFeatures-1}\SlicedParameterLayer_{c' + (\NFullresFeatures+1) c}[x,y] \,
\FullresFeatures_{c'}[x,y]
\label{eq:affine_model} 
\end{equation}}

\begin{figure}[!t] 
    \centering
    \includegraphics[width=\columnwidth]{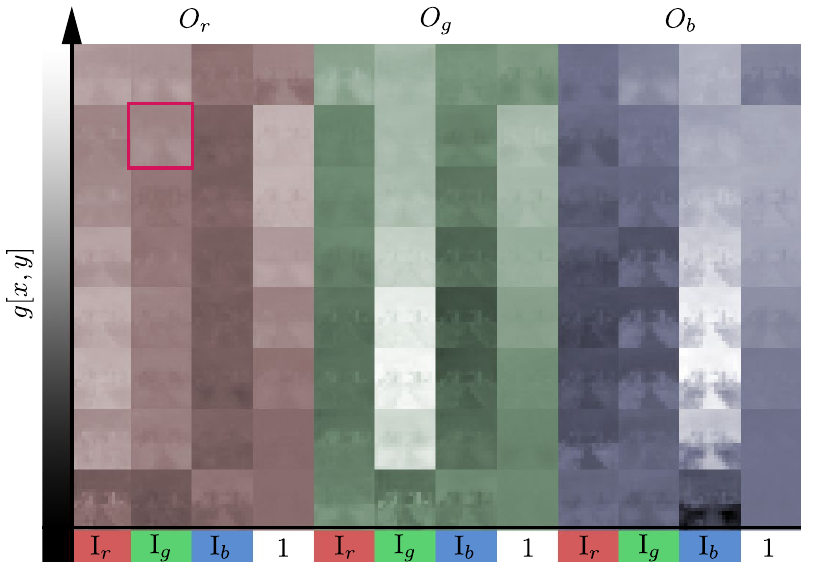}
    \caption{Coefficient maps for the affine color transform. The vertical axis
    corresponds to the learned guidance channel, while the horizontal axis
    unrolls the 3x4 sets of coefficients. Each thumbnail, one example of which
    is highlighted, shows	a 16x16 low-resolution map. }
    \label{fig:coefficient_map}
\end{figure}

Interpolated affine transformations similar to this have been used successfully
for matting~\cite{levin2008_closed}, intrinsic image
decomposition~\cite{bousseau2009_user} and time of day
transfer~\cite{shih2013_data}.
For such models, the size of the patch in which the affine model is fit
drives the trade-off between efficiency and quality. 
At the extreme, it is always possible to achieve a perfect reconstruction of
any operator by fitting an independent model at every pixel (\ie, the patch
size is $1\times 1$).
For small patches (\eg, $3\times 3$), an affine model can faithfully
reproduce many image operators. 
As the patch grows larger, the affine relationship no longer holds for all but
trivial operators, though others have shown that this limitation can be
mitigated using piecewise linear functions~\cite{yuan2011_localpiecewiselinear}
or non-linear and edge-aware components~\cite{Gharbi2015_recipe}.
See Figure~\ref{fig:coefficient_map} for a visualization of the 3D bilateral
grid of affine coefficients \ParameterLayer corresponding to the input/output
pair in Figure~\ref{fig:network}. One of the 12 channels of the 2D
coefficients after slicing can also be seen in Figure~\ref{fig:network}.

\subsection{Training procedure}
\label{sec:training}

We train our network on a dataset $\dataset=\left\{ \left( \Input_i, \Output_i
\right) \right\}_i$ of full-resolution input/output pairs for a given operator. We optimize the
weights and biases by minimizing the $L_2$ loss on this training set:
\begin{equation}
    \Loss=\frac{1}{\abs{\dataset}}\sum_i \norm{\Input_i-\Output_i}^2
    \label{eq:loss}
\end{equation}
We additionally regularize the weights with an $L_2$ weight decay of \weightdecay. 
The weights for the convolutional and fully-connected layers are initialized
according to \cite{He2015_weight_init} and the biases are initialized to 0.
We use batch normalization \cite{ioffe2015_batch} between each pair of
intermediate feature maps, and we optimize the network parameters with the ADAM
solver \cite{Kingma2014_adam_solver}.
We train with a batch size of 4 to 16 (depending on the resolution) and a
learning rate of \learningrate. The remaining parameters in ADAM are kept
to the values recommended by the authors.
Our model is implemented in Tensorflow \cite{tensorflow2015-whitepaper} and
Halide \cite{JRK2013_halide}. For all experiments, models are trained on an
NVIDIA Titan X (Maxwell) for 30 epochs, which typically takes 2--3 days.


\section{Results}

\ADD{We evaluate our model's ability to reproduce both algorithmic image
operators (Section~\ref{sec:operators_results}) and human-annotated retouches
(Section~\ref{sec:human_results}).}
\ADD{Our model is faster than both standard neural networks and
state-of-the-art filter approximation techniques and runs in real-time on mobile
device (Section~\ref{sec:performances})}.

\begin{figure*}[!p]
    \includegraphics[width=0.98\textwidth]{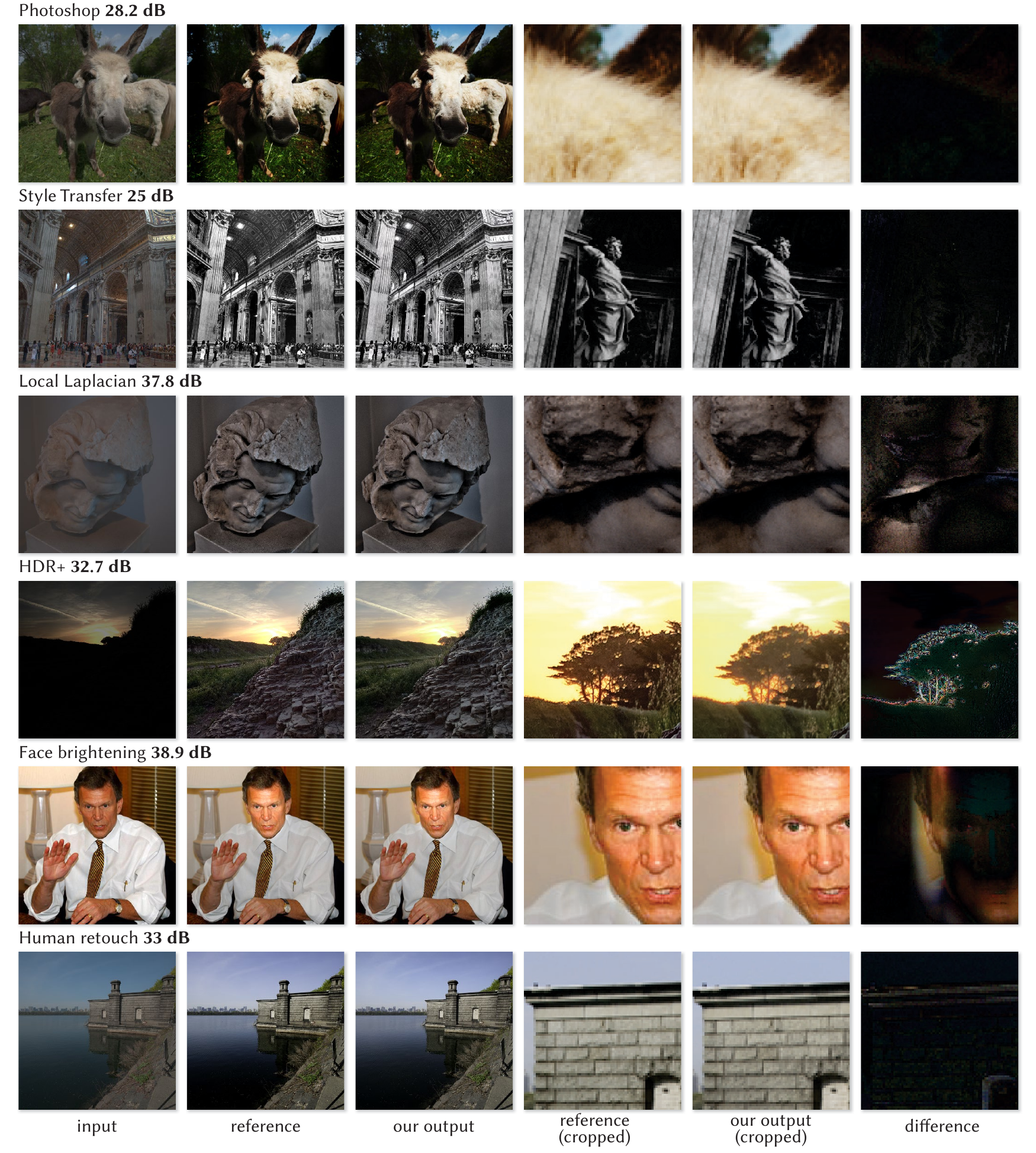}
    \caption{
	Our method can learn accurate and fast approximations of a wide
	variety of image operators, by training on input/output pairs processed by
	that operator.
	These operators can be complicated ``black box'' image processing pipelines
	where only a binary is available, such as HDR+ or Photoshop filters/actions.
	Some operators, such as face-brightening, requires semantic understanding.
	Our model is even capable of learning from highly subjective
	human-annotated input/output pairs, using the MIT-Adobe FiveK
	dataset.
  \ADD{The difference is rescaled to use the full $[0,1]$ range.}
	}
  \label{fig:gallery}
\end{figure*}

A selection of our results on different tasks can be seen in Figure~\ref{fig:gallery}.
Our output is generally accurate and, even when it differs from the
ground-truth, it remains plausible.
Despite the heavy spatial and bilateral downsampling inherent to our approach,
image artifacts are rare and unobjectionable.
This is because of the edge-aware nature of the bilateral grid and our model's
capacity to learn smooth output transformations.
Our outputs are usually slightly softer (\eg on the HDR+ example of
Figure~\ref{fig:gallery}) because the highest-frequency transformations
like sharpening and the correction of chroma aberrations can introduce new edges
not present in the input, which our model does not handle.
\RM{Please see the supplemental material for additional results.}

\subsection{\ADD{Reproducing image operators}}
\label{sec:operators_results}

We evaluate the accuracy of our model on several tasks composed of
programmatically-defined image operators:
\begin{itemize}[.]
\item \emph{HDR+}~\cite{hasinoff2016_burst} -- a complex
hand-engineered photographic pipeline that includes color correction,
auto-exposure, dehazing, and tone-mapping.
\item the \emph{Local Laplacian} filter~\cite{paris2011_local} -- an edge-preserving,
multi-scale (yet non-scale-invariant) operator used for detail enhancement (we
use two different strengths for the effect),
\item the \emph{Style Transfer} task of \cite{aubry2014_fast} (which happens to be
based on the Local Laplacian),
\item a \emph{Face brightening} task using a dataset of labeled faces \cite{fddbTech},
\item several different black-box \emph{Adobe Photoshop (PS)} filters and user-created
``actions''\footnote{\url{http://designbump.com/photoshop-actions-for-instagram-effects/}}.
\end{itemize}
PSNRs for these tasks using our model and baseline approaches can be found in
Table~\ref{tab:benchmark}.

We use two variants of the style transfer task.
In the first variant (\emph{Style Transfer}), we learn to transform any new
input towards a unique fixed style.
In the second, more challenging variant (\emph{$n$-Styles Transfer}) we adapt
our network to take two input images (concatenated along their channel axis) and
predict the results of transferring the style of one image to the other (again using the
algorithm of Aubry~et~al.~\shortcite{aubry2014_fast}).
In this variant the network does not learn to predict a single consistent
output; but rather, it learns to extract the desired transformation from the
target image and apply that transformation to the input image.

\begin{table}[!b]
\caption{
  We compare accuracy to Bilateral Guided Upsampling (BGU) and Transform Recipes
  (TR). Note that BGU and TR are ``oracle'' techniques, as they run the code used to
  evaluate each image operator at a reduced or full resolution, and so can be
  thought of as providing an upper-bound on performance. Despite its
  disadvantage, our model sometimes performs better than these oracle baselines
  due its expressive power and ability to model non-scale-invariant operators.
}
\label{tab:benchmark}
\centering
\begin{tabular}{lccc}
Task (PSNR, dB) & Ours & BGU & TR \\
\hline
HDR+                              & 28.8  & 26.9  & 29.0  \\
Local Laplacian                   & 33.5  & 32.2  & 38.6  \\
Local Laplacian (strong)          & 30.3  & 20.6  & 31.8  \\
Face brightening                  & 33.7  & 30.9  & 33.9  \\
Style Transfer                    & 23.9  & 21.9  & 31.7  \\
$n$-Styles Transfer               & 27.6  & 21.9  & 33.7  \\
PS eboye                          & 45.0  & 33.5  & 41.5  \\
PS early bird                     & 25.9  & 22.2  & 32.8  \\
PS instagram                      & 40.3  & 37.1  & 40.7  \\
PS infrared                       & 38.4  & 34.5  & 38.7  \\
PS false colors                   & 38.1  & 34.3  & 38.6  \\
PS lomo-fi                        & 26.2  & 24.1  & 34.4  \\
\end{tabular}
\end{table}

\begin{figure}[!t]
    \centering
    \includegraphics[width=0.9\columnwidth]{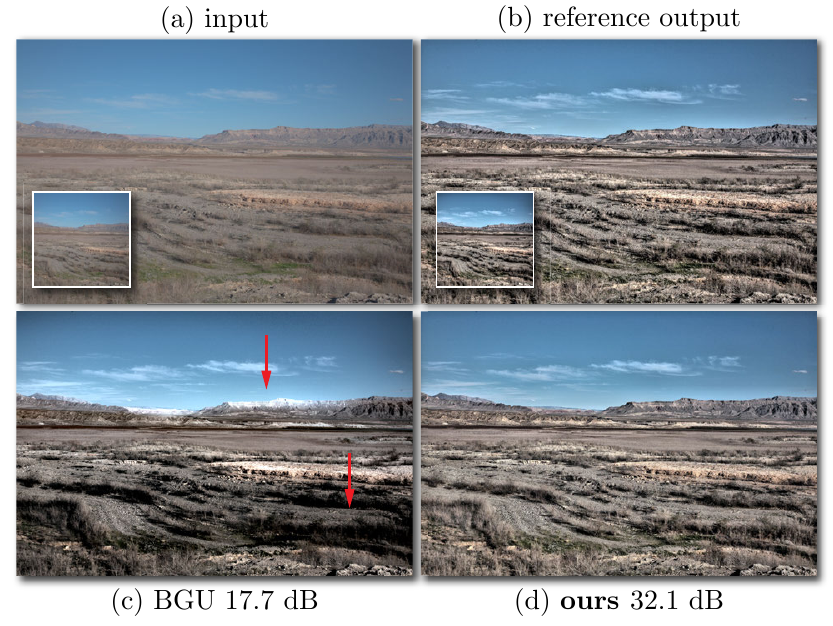}
    \caption{
	  Our method (d) can learn to replicate the correct effect (b) for
	  operations that are not scale invariant, such as the Local Laplacian filter
	  shown here (a--b).
	  Methods like Bilateral Guided Upsampling that only apply the operation at
	  low-resolution (insets (a--b)) produce a different-looking output (c).
	  The difference is most noticeable in the areas pointed by the arrows.
	  }
    \label{fig:non_scale_invariant}
\end{figure}

\subsubsection{\ADD{Datasets}}
Besides HDR+ and the face brightening dataset, all the effects were applied to
the \emph{unprocessed} set of the MIT ``FiveK'' dataset~\cite{Bychkovsky2011_mit5k}. We
reserve 500 images for validation and testing, and train on the remaining 4500.
We augment the data with random crops, flips and rotations.
We generated the dataset for \emph{$n$-Styles Transfer} by mapping each image in the
MIT ``FiveK'' dataset to 100 distinct images (the style targets) .

\subsubsection{\ADD{Baseline}}
The previous work closest in spirit to our goals are Bilateral Guided Upsampling
(BGU)~\cite{Chen2016_bgu} and Transform Recipes (TR)~\cite{Gharbi2015_recipe} to
which we compare our outputs.
However, whereas our technique learns a photographic operator offline from a
dataset of images, BGU and TR use no prior training and instead fit
specially-tailored models to an input/output pair in an online fashion.
BGU and TR therefore require direct access to the image operator, as they
require the ability to run that image operator on images (either downsampled
on-device or full-resolution on a server, respectively).
This makes our comparisons against these baselines somewhat biased against our
technique, as these baselines make more limiting assumptions about what is
available, and also cannot learn to approximate a general instance of an
image operator from data.
Regardless, we report metrics for these techniques as a kind of ``oracle''
baseline.

Transform Recipes assumes that a mobile device would send a highly compressed
(and therefore degraded) image to a server for processing, and would recieve an
inexpensive ``recipe'' for approximating an image transformation from that
server.
Because TR's client-server setup is not relevant to the scope of this paper, we
run the model (using the authors' recommended settings) on uncompressed,
full-resolution images, thereby improving output quality and making our TR
baseline as competitive as possible. In the intended use case of the method, the
image quality typically decreases by 3--5 dB.

BGU assumes that the image operator be run on a low-resolution version
of the input before fitting the model to the low-res input/output pair.
We could not run the  HDR+ filter at low resolution, so we used full-resolution
input/output pairs and created the low-resolution inputs to BGU by downsampling.
We do however follow the correct procedure for the \emph{Local Laplacian} and
\emph{Style Transfer} tasks for which we have an implementation and directly
apply the filter at low resolution.
For these non scale-invariant tasks, the advantage of our technique becomes
clearer (Figure~\ref{fig:non_scale_invariant}).

\subsection{\ADD{Learning from human annotations}}
\label{sec:human_results}

\begin{table}[!t]
\caption{Mean $L_2$ error in L*a*b* space for retouches from the 5 photographers
  in the MIT5k dataset (A,B,C,D,E); lower is better. Our algorithm is capable of
  learning a photographer's retouching style better than previous work, yet runs
  orders of magnitudes faster. The comparisons in the first two groups are
  evaluated on the dataset from photographer C favored by previous techniques;
  see main text for details. In the third group we report our results on the
  remaining 4 photographers for completeness. Metrics taken from previous work
  \protect\cite{Yan2016_automatic,hwang2012_context} are denoted by $^\dagger$.}
\label{tab:mit_experts}
\def\sym#1{\ifmmode^{#1}\else\(^{#1}\)\fi}
\centering
\begin{tabulary} {1\linewidth} {c l *{2}{S[table-format=2.2,detect-weight,table-align-text-post=false]}}
photographer & method  & {\lab}        & {L-only} \\
\midrule
\multirow{4}{*}{\shortstack{ C\\\textit{random250}}} & ours                        & \bfseries 7.8 & \bfseries 5.5 \\
                                                    & Yan~\shortcite{Yan2016_automatic}    & 9.85$^\dagger$          & \na \\
                                                    & Bychkovsky~\shortcite{Bychkovsky2011_mit5k} & \na           & 5.8 \\
                                                    & Hwang~\shortcite{hwang2012_context}    & 15.01$^\dagger$        & \na \\
\midrule
\multirow{4}{*}{\shortstack{ C\\\textit{highvar50}}} & ours                        & \bfseries 7.1 & \bfseries 5.2 \\
                                                    & Yan~\shortcite{Yan2016_automatic}    & 8.36$^\dagger$          & \na \\
                                                    & Hwang~\shortcite{hwang2012_context}    & 12.03$^\dagger$        & \na \\
\midrule
 A & ours                        & 11.7          & 9.8 \\
 B & ours                       & 7.4           & 5.0 \\
 D & ours                       & 10.0          & 7.7 \\
 E & ours                        & 8.8           & 6.2 \\
\end{tabulary}
\end{table}

We also evaluate accuracy with regards to human annotations using the MIT-Adobe
``FiveK'' dataset~\cite{Bychkovsky2011_mit5k}, and our performance compared to
previous work is presented in Table~\ref{tab:mit_experts}.
This task measures our model's ability to learn a highly subjective image operator
which requires a significant amount of learning and semantic reasoning.
We report mean $L_2$ error in L*a*b* space (lower is better) for retouches by the
5 photographers (A,B,C,D,E) in the MIT ``FiveK'' dataset, though previous work only
presents results on photographer C~\cite{Yan2016_automatic,hwang2012_context}.
We use the ``Random 250'' and ``High Variance 50'' dataset splits presented in
\cite{hwang2012_context}, which have 250 randomly-chosen and 50 user-weighted
images in the test set, respectively.

This is a much more difficult task, and inconsistencies in the retouches of
photographers has been pointed out previously~\cite{Yan2016_automatic}. For
example we found that retoucher B in this dataset was more self-consistent, and
was easier for our network to learn.
\RM{Nonetheless, our model outperforms previous work and predicts reasonable image
corrections, distinct for each retoucher we trained it on.}
\ADD{Nonetheless, our model, trained separately on each artist's corrections,
consistently predicts reasonable adjustments and outperforms previous work.}

\begin{figure}[!t]
    \centering
    \includegraphics[width=\columnwidth]{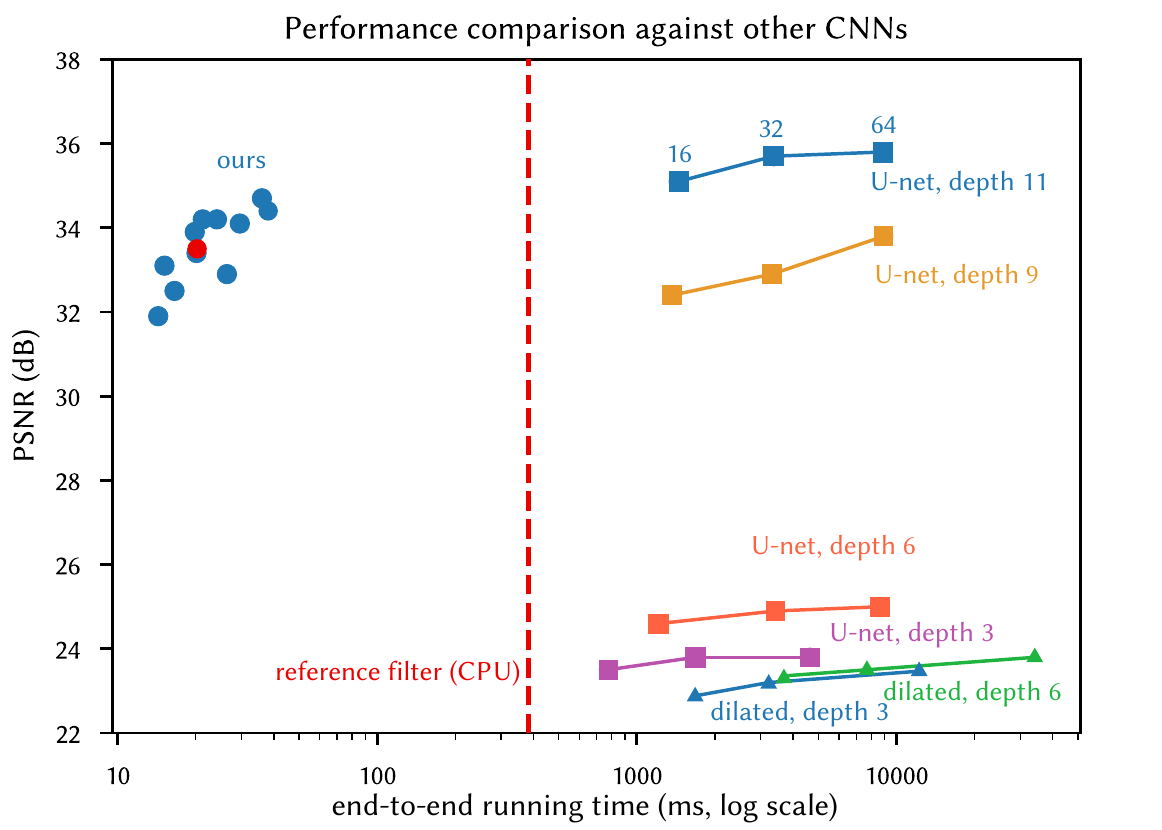}
    \caption{\ADD{We compare the speed and quality of our algorithm against two
    modern network architectures: \emph{U-Net} (adapted from \cite{isola2016})
    and \emph{dilated convolutions} \cite{yu2015_dilated}.
    The runtimes were averaged over 20 iterations, processing a 4 megapixel
    image on a desktop CPU. The PSNR numbers refer to the \emph{Local Laplacian}
    task.
    Given an insufficient \emph{depth}, U-Net and dilated convolutions fail to
    capture the large scale effects of the Local Laplacian filter, leading to
    low PSNRs. Competitive architectures run over 100 times slower than ours,
    and use orders of magnitude more memory. 
    Our model's performance is displayed for a range of parameters. 
    The version we used to produce all the results is highlighted in red.
    See Figure~\ref{fig:grid_search}
    for details on the speed/quality trade-off of our model. 
    }}
    \label{fig:performance_comparison}
\end{figure}

\begin{figure}[!t]
    \centering
    \includegraphics[width=.8\columnwidth]{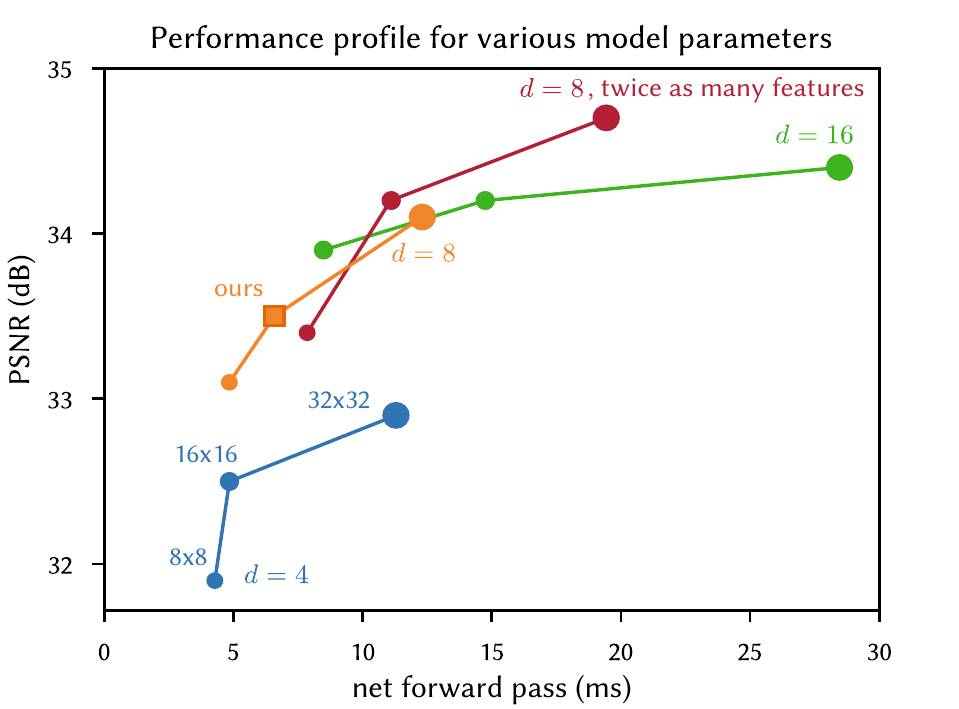}
    \caption{\ADD{We show PSNRs for the \emph{Local Laplacian} task and the
    computation time required to predict the bilateral coefficients with several
    settings of our model's parameters. 
    Each curve represent a grid depth \GridDepth. For each curve the grid's spatial
    resolution varies in $\{8, 16, 32\}$. 
    The reference model we used to produced all the results is highlighted with
    a square marker.
    Unsurprisingly, models with larger grid depth perform better (green). 
    Doubling the number of intermediate features also provides a 0.5 dB
    improvement (red curve). Runtimes were measured on an Intel Core
    i7-5930K.}}
    \label{fig:grid_search}
\end{figure}

\subsection{Performance}
\label{sec:performances}

We implemented our technique on a Google Pixel phone running Android 7.1.1.
Our implementation processes viewfinder-resolution $1920\times1080$ images in
realtime, at $40$--$50$\hz.
We extract 8-bit preview frames in YUV420 format using the \emph{Camera2} API.
These images are downsampled to $256\times 256$, converted to floating point RGB, then fed into our network.
After the network produces its output (a bilateral grid of affine coefficients),
we transfer them to the GPU as a set of three 3D RGBA textures, where they are sliced
and applied to the full-resolution input to render the final processed preview.
Overall throughput is under $20$\ms, with $14$\ms spent on inference (CPU),
overlapped with $1$\ms to upload coefficients and $18$\ms to render on the GPU.
As a point of comparison, running an optimized implementation \cite{JRK2013_halide}
of the Local Laplacian filter~\cite{paris2011_local} on the same device takes over
$200$\ms.
Running the same filter at the reduced $256\times 256$ resolution and applying
Bilateral Guided Upsampling~\cite{Chen2016_bgu} with the same grid dimensions
takes $17$\ms (compared to our $14$\ms) but loses some of the filter's
intended effect (Figure~\ref{fig:non_scale_invariant}).
Our processing time scales linearly with input size, taking $61$\ms to process a
12-megapixel image.
While it usually has higher fidelity, Transform Recipes~\cite{Gharbi2015_recipe}
requires $2.95$ seconds per image, nearly two orders of magnitude below
real-time viewfinder performance.
Mosty notably, neither Transform Recipes nor Bilateral Guided Upsampling can
apply effects learned from human retouches, or ``black box'' operators such as Photoshop filters or HDR+.

Other recent neural-network based architectures that could be used for such
learning are also far from real-time.
\ADD{In Figure~\ref{fig:performance_comparison}, we compare our technique
  against a U-Net architecture \cite{ronneberger2015_unet} adapted from Isola~et~al.~\shortcite{isola2016},
  and a linear network based on dilated convolutions~\cite{yu2015_dilated}. We
  explore several settings for the \emph{depth} (number of layers, 3 to 11) and the
  \emph{width} (number of filters, 16 to 64) in these architectures, covering a variety of speed and quality
  levels. For U-Net, ``depth'' refers to the number of downsampling steps and
  ``width'' refers to the channels in the first convolutional layers (these are doubled
  at each downsampling step, see Isola~et~al.~\cite{isola2016} for details).
  In the dilated convolution network, ``depth'' is the number of dilated
  convolution layers, and ``width'', the number of channels in each layer.
  Our hybrid CPU/OpenGL technique is over 2 orders of magnitude faster
  than both architectures on a desktop CPU. On GPU (not shown), the performance gap is
  identical for the forward pass of the network, but data transfer becomes the
  bottleneck for our method. End-to-end, our runtime is still over an order of
  magnitude faster. Moreover, both U-Net and dilated convolution require
  significantly more memory, which makes them ill-suited for mobile processing.
  For this benchmark we used an Intel Core i7-5930K at 3.5GHz with 4 cores and a Titan X
  (Maxwell) GPU.}

  \ADD{
  We explored the speed/quality trade-offs of our architecture for the \emph{Local
  Laplacian} task varying several parameters: changing the depth of the grid
  \GridDepth from 4 to 16, the grid's spatial dimensions from $8\times 8$ to
  $32\times 32$ and doubling the number of channels (compared to the numbers
  reported in Table~\ref{tab:architecture}). The summary can be found in
  Figure~\ref{fig:grid_search}.}
\RM{requiring $2.13$
or $2.36$ seconds respectively to process images of our size on high-end desktop
GPUs (according to the throughputs reported in each paper)}


\subsection{\ADD{Discussion and limitations}}

\begin{figure}[!tb]
    \centering
    \includegraphics[width=.82\columnwidth]{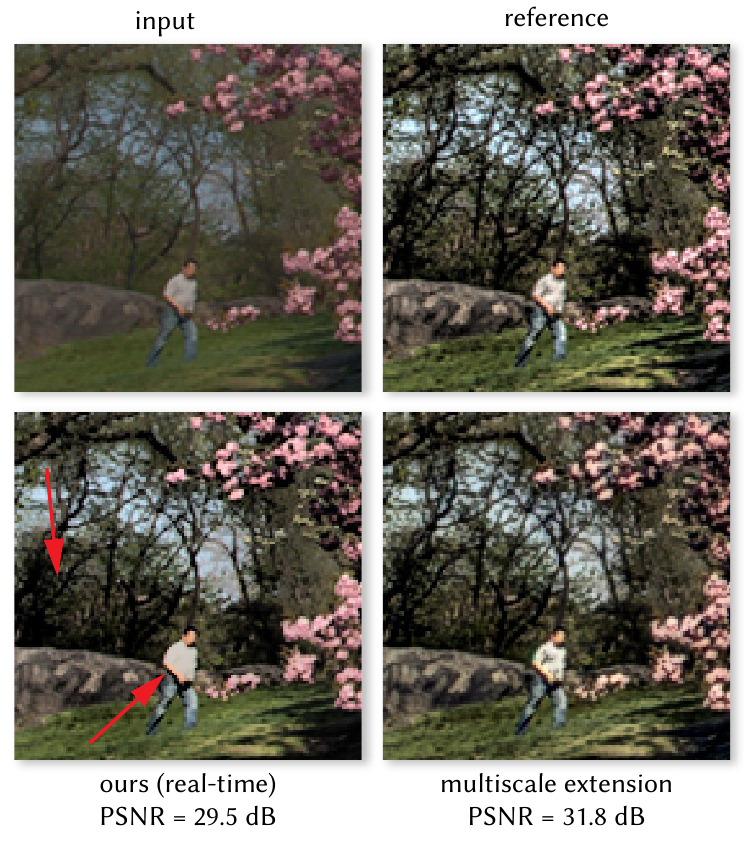}
    \caption{\ADD{At the expense of extra computation at full-resolution,
        our model can be extended with richer affine regression features.
        Here, by using a 3-level Gaussian pyramid as features
        \FullresFeatures, we can better capture the high-frequency
        details in the the \emph{Local Laplacian (strong)} task.}}
    \label{fig:multiscale}
\end{figure}

\ADD{
All our results use the simplest full-resolution features
$\FullresFeatures=\Input$; \ie, both the guide \Guide and the affine regression
targets are the color channels of the input image (Section~\ref{sec:fullres}).
If one relaxes the real-time rendering constraint, one can extend our model by
extracting features from the high-resolution image.
In Figure~\ref{fig:multiscale}, we show an example where \FullresFeatures is a 
3-level Gaussian pyramid. The bilateral grid then contains $3\times 12 = 36$
affine parameters (12 for each scale). Accordingly we triple the number of
intermediate features in the network compared to the numbers in
Table~\ref{tab:architecture}. This roughly slows down the network by a factor 3-4,
but provides a 2 dB boost in quality on the \emph{Local Laplacian (strong)} task.}

\begin{figure}[!t]
    \centering
    \includegraphics[width=.95\columnwidth]{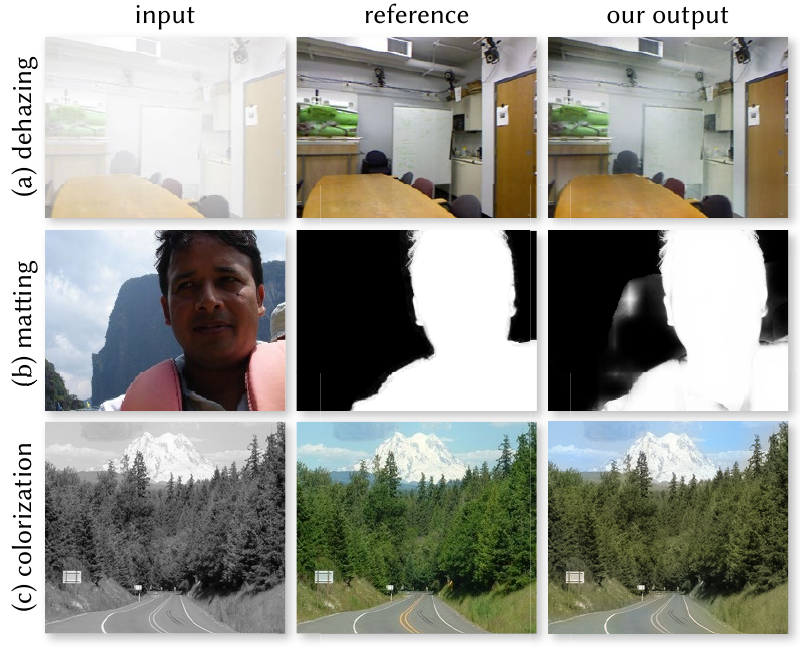}
    \caption{\ADD{Our algorithm fails when the image operator strongly violates our modeling
    assumptions. 
    (a) Haze reduces local contrast, which limits the
    usefulness of our guidance map. It also destroys image details
    that cannot be recovered with our affine model (\eg, on the whiteboard).
    (b) Matting has successfully been modeled by locally affine models on
    $3\times 3$ neighborhoods~\cite{levin2008_closed}. However, this affine
    relationship breaks down at larger scales (like a grid cell in our model)
    where the matte no longer follows tonal or color variations and is mostly
    binary. This limits the usefulness of our bilateral grid.
    (c) For colorization, the learned guidance map is at best a nonlinear
    remapping of the grayscale input. Our model can thus only learn a local
    color per discrete intensity level, at a spatial resolution dictated by the grid's
    resolution. Our output is plagued with coarse variations of colors that are muted
    due to our $L_2$ loss (see the road line, and the tree/sky boundary).
}}
    \label{fig:failures}
\end{figure}

We \ADD{also} explored using our architecture to learn tasks beyond image
enhancement, like matting, colorization, dehazing, and monocular depth
prediction. 
These experiments had limited success, as the strong modeling assumptions
required for fast photographic correction make our model poorly suited to
different tasks whose output cannot be easily expressed as local pointwise
transformations of the input image \ADD{(Figure~\ref{fig:failures})}.

\section{Conclusion}

We have introduced a new neural network architecture that can perform image
enhancement in real-time on full-resolution images while still capturing
high-frequency effects.
Our model is trained using pairs of input/output images, allowing it to learn
from a reference implementation of some algorithm or from human adjustments.
By performing most of its computation within a bilateral grid and by
predicting local affine color transforms, our model is able to strike the right
balance between expressivity and speed.
To build this model we have introduced two new layers: a data-dependent
lookup that enables slicing into the bilateral grid, and a multiplicative
operation for affine transformation.
By training in an end-to-end fashion and optimizing our loss function at full
resolution (despite most of our network being at a heavily reduced resolution),
our model is capable of learning full-resolution and non-scale-invariant
effects.
The accuracy of our model has been demonstrated on a variety of different image
operators, pipelines, and subjective human-annotated datasets.

\begin{acks}
We thank the SIGGRAPH reviewers for their constructive comments.
Special thanks to Marc Levoy for his valuable feedback.
This work was partially funded by Toyota.
\end{acks}

\bibliography{_hdrnet}

\end{document}